\documentstyle[11pt,newpasp,twoside]{article}
\def\edcomment#1{\iffalse\marginpar{\raggedright\sl#1\/}\else\relax\fi}
\reversemarginpar

\begin{document}
\title{Interactions of High- and Low-Mass Planets with Protoplanetary Disks}
  \author{Stephen~H. Lubow}
\affil{Space Telescope Science Institute, 3700 San Martin Drive,
Baltimore, MD 21218}
\author{Matthew~R. Bate}
\affil{School of Physics, University of Exeter, Exeter EX4 4QL, UK\\}
 \author{Gordon~I. Ogilvie}
\affil{Institute of Astronomy, University of Cambridge,
Madingley Road, Cambridge CB3 0HA, UK}

\begin{abstract}
We present three-dimensional numerical simulations 
of the interaction of a circular-orbit planet with a protoplanetary
disk.  We calculate the flow
pattern, the accretion rate, and torques on a planet. We consider
planet masses ranging from 1 Earth mass to 1 Jupiter mass.

\end{abstract}

\section{Introduction}
Planet formation is believed to involve accretion
from the surrounding disk of material. We consider this gas
accretion onto a planet. A non-Keplerian
flow pattern develops near the planet. Furthermore,
the planet is subject to gravitational torques due to its interaction with the
gas disk which result in planetary migration. We have
extended the 2D simulations by Lubow, Seibert, and Artymowicz (1999)
to 3D, in order to analyze these effects more realistically.
We expect 3D effects to be important for planet masses below $1$M$_{\rm J}$,
since the Hill sphere radius becomes smaller than the disk thickness.

\section{Computational Procedure}
The calculations were performed by using the 3D, spherical
version of the ZEUS finite-difference
code (Stone et al. 1992). The disk model is similar
to that of Lubow et al. (1999), except that 
we include the third dimension. The disk is locally
isothermal with $T(r) \propto r^{-1}$, and disk thickness ratio
$H/r=0.05$. The initial surface density is taken to be
$\Sigma(r) \propto r^{-1/2}$ away from the planet.
The total disk mass is assumed to be 0.0075 solar masses
in the radial interval from 1.6 to 21 AU. The planet's orbital
radius is taken to be 5.2 AU.  Thus, the surface density of the
unperturbed disk at the planet's orbital radius is 75 g cm$^{-2}$.
We apply a constant kinematic viscosity
$\nu = 10^{-5}$ which, at the planet's radius, corresponds
to viscosity parameter $\alpha =  4 \times 10^{-3}$.

We adopt a grid $(r, \theta, \phi) = (180,36,288)$ that is
non-uniform in $r$ and $\phi$. The highest resolution is achieved
close to the planet. Zones near the planet measure 0.002875 in $r$
and $\phi$, and twice this value in $\theta$. We model 4 vertical scale
heights above the midplane.

We consider separately planets of masses 1, 0.1, 0.03, 0.01, and 
0.003 M$_{\rm J}$ which orbit about a 1 $M_{\odot}$ star. The latter planet 
mass corresponds to 1 Earth
mass. In all cases, the flow is resolved within the planet's Hill sphere.

\section{Surface Density}
The planet-disk interaction modifies the disk in the vicinity of
the planet (Fig.~1). In Fig.~2, we plot the azimuthally-averaged 
disk surface density. Only a 1$M_{\rm J}$ planet opens
a deep gap and a mass of at least 0.1$M_{\rm J}$ is required for
the surface density to be perturbed significantly. 
Lower mass planets result in much less severe surface-density perturbations.
\begin{figure}
\plotfiddle{lubow1a.eps}{2.5cm}{0}{32}{32}{-195}{-128}
\plotfiddle{lubow1b.eps}{2.5cm}{0}{32}{32}{0}{-44}
\plotfiddle{lubow1c.eps}{2.5cm}{0}{32}{32}{-195}{-143}
\plotfiddle{lubow1d.eps}{2.5cm}{0}{32}{32}{0}{-59}
\caption{Density and streamlines at the disk midplane for 1.0, 0.1, 0.03, 
and 0.003 M$_{\rm J}$ planets.}
\end{figure}
\begin{figure}
\plotfiddle{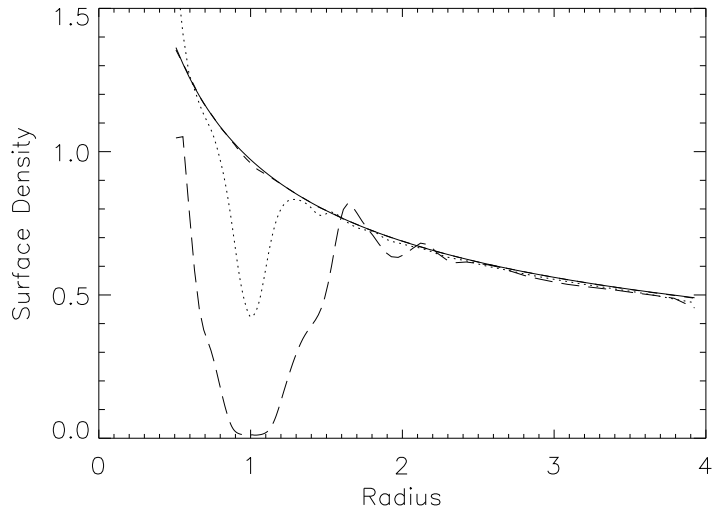}{6.0cm}{0}{60}{60}{-180}{-230}
\caption{Azimuthally-averaged disk surface density for 1.0 (long-dashed), 0.1 (dotted), 0.03 (short-dashed), and 0.01 (solid) M$_{\rm J}$ planets.  Only planets with masses $\geq 0.1$ M$_{\rm J}$ produce significant perturbations.}
\end{figure}

\section{Migration Rates}

In Fig.~3, left panel, we plot the migration rates of the 5 planets.
We find that the migration rate increases linearly with planet mass,
as expected for Type I migration where the planet
does not open a gap in the disk.  We also plot the
migration rates expected from the linear theories of 
Ward (1997) and Tanaka et al.~(2002).  Ward's calculations approximate
the disk as two-dimensional and do not include co-rotation resonances.
Tanaka et al.~consider three-dimensional disks and both Lindblad and
co-rotation resonances.  Our results are in excellent agreement with 
Tanaka et al.\ for $\leq 0.1$ Jupiter masses (i.e.\ in the Type I regime).
\begin{figure}
\plotfiddle{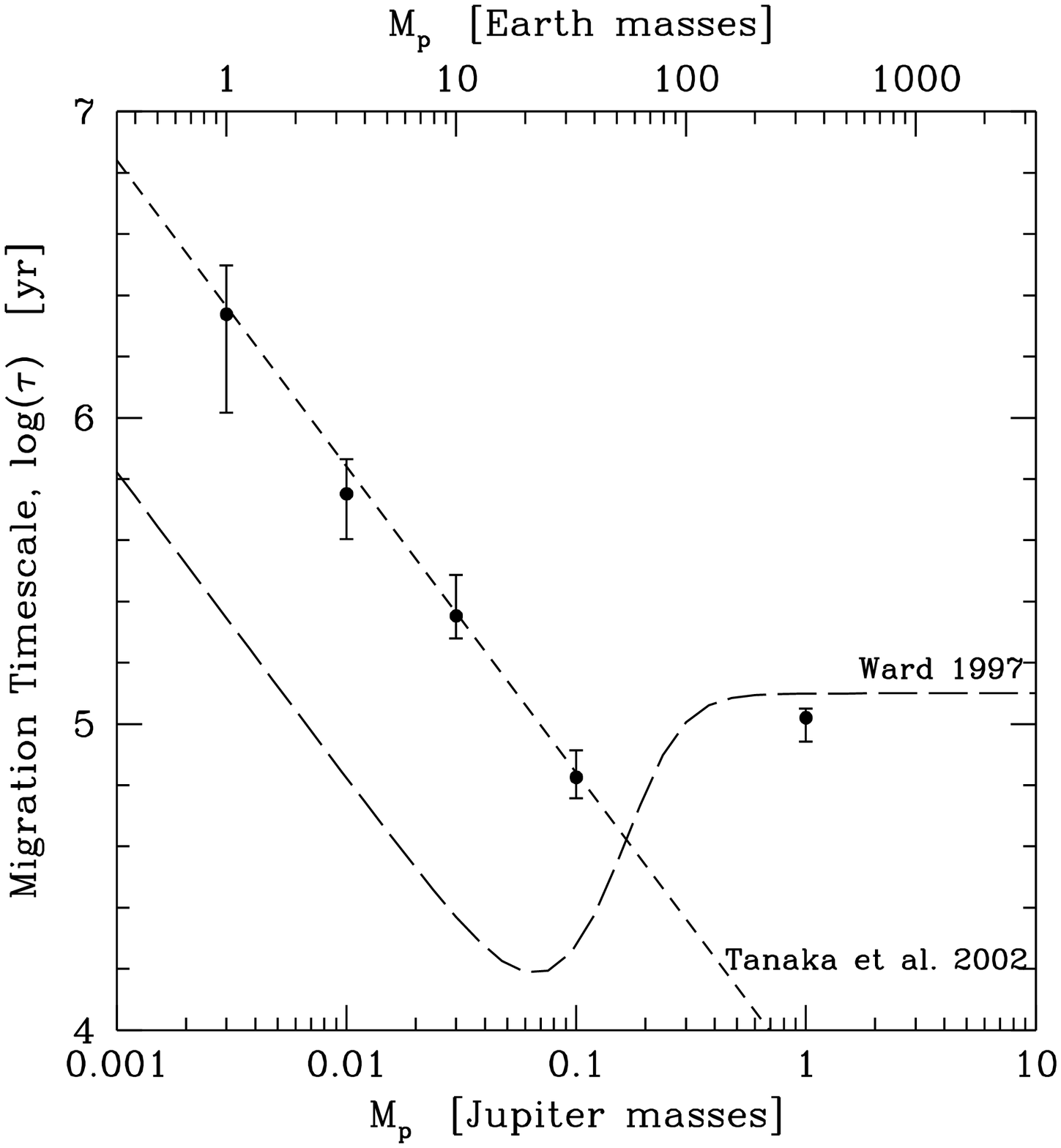}{2.5cm}{0}{32}{32}{-195}{-138}
\plotfiddle{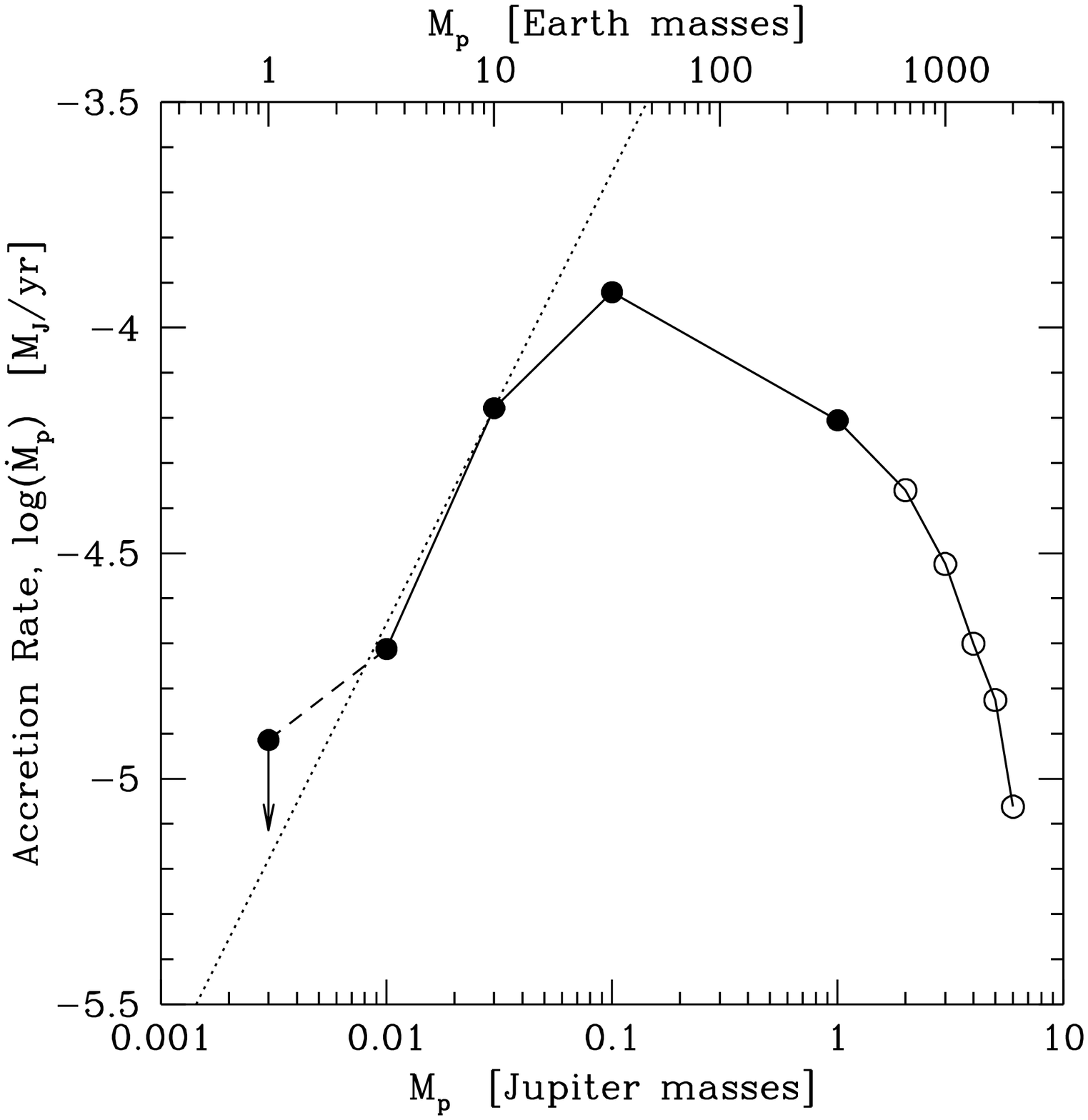}{2.5cm}{0}{32}{32}{0}{-54}
\caption{Migration time, $\tau$, and accretion rate, $\dot M_{\rm p}$, 
versus planet mass.  Our results are plotted using solid circles.  
Open circles give accretion rates for very high-mass planets 
taken from Lubow et al.\ (1999).
The long and short dashed lines give the linear predictions of migration 
rates from Ward (1997) and Tanaka et al.\ (2002), respectively.  
We find the accretion rates of low-mass planets increase linearly with
mass (dotted line) 
until $\approx 0.1$ M$_{\rm J}$ and decrease for higher masses.
}
\end{figure}

\section{Accretion Rates}
In Fig.~3, right panel, we plot the 
accretion rates onto the planet against planet mass
for the 5 planets (filled circles). Also plotted are the rates obtained
for higher mass planets from the two-dimensional calculations of
Lubow et al (1999).

\section{Conclusions}
\begin{itemize}
\item The migration rate scales linearly with $M_{\rm p}$ for planets 
that do not open a gap.  The results are in excellent agreement with
the linear analysis of Tanaka et al.~(2002).  Planets with masses 
$>0.1$ M$_{\rm J}$ migrate on the disc's viscous timescale.
\item The mass accretion rate increases linearly with planet mass $M_{\rm p}$ 
for low-mass planets and peaks at $\approx 0.1$ M$_{\rm J}$. At higher masses, the mass
accretion rate declines as a gap is opened in this disk.
\end{itemize}

\acknowledgments
The computations reported here were performed using the UK Astrophysical 
Fluids Facility (UKAFF).
SHL acknowledges support from NASA grants NAG5-4310 and NAG5-10308.

\end{document}